\documentclass[prd,preprint,nofootinbib,superscriptaddress]{revtex4}
\usepackage{amsmath,amssymb}
\usepackage{cancel}
\usepackage{graphicx}

\begin{document}

\title{Phenomenology of Non-Custodial Warped Models}
\date{\today}
\preprint{CAFPE-159/11}
\preprint{UGFT-289/11}

\author{Adri\'an Carmona}
\affiliation{
CAFPE and Departamento de F\'{\i}sica Te\'orica y del 
Cosmos, \\
Universidad de Granada, E-18071 Granada, Spain}
\author{Eduardo Pont\'on}
\affiliation{Department of Physics, Columbia University,
538 W. 120th St, New York, NY 10027, USA}
\author{Jos\'e Santiago}
\affiliation{
CAFPE and Departamento de F\'{\i}sica Te\'orica y del 
Cosmos, \\
Universidad de Granada, E-18071 Granada, Spain}

\begin{abstract}
We study the effect of bulk fermions on electroweak precision
observables in a recently proposed model with warped extra dimensions
and no custodial symmetry.  We find that the top-quark mass, together
with the corrections to the $Zb_{L}\bar{b}_{L}$ vertex and the
one-loop contribution to the $T$ parameter, which is finite, impose
important constraints that single out a well defined region of
parameter space.  New massive vector bosons can be as light as $\sim
1.5$ TeV and have large couplings to the $t_R$ quark, and suppressed
couplings to the $t_L$, $b_L$ and lighter quarks.  We discuss the
implications for searches of models with warped extra dimensions at
the LHC.
\end{abstract}

\pacs{}

\rule{0mm}{1.5cm}

\maketitle

\section{Introduction}

Models with warped extra dimensions~\cite{Randall:1999ee} are
excellent candidates for physics beyond the Standard Model.  They
solve the hierarchy problem, provide a rationale for the observed
flavor structure, and can potentially give striking signatures at the
LHC. Besides, they are weakly coupled duals of strongly coupled
four-dimensional (conformal) theories and can therefore provide an
intuition and quantitative predictions for models of strong
electroweak symmetry breaking (EWSB).

Models based on a purely AdS$_5$ background and a minimal field
content are strongly constrained by electroweak precision tests
(EWPT).  In particular, large contributions to the Peskin-Takeuchi $T$
parameter~\cite{Peskin:1991sw} can easily push the scale of new
physics beyond the reach of the LHC~\cite{Huber:2001gw}.  The large
top mass requires the third generation $SU(2)_{L}$ quark doublet (as
well as the top singlet) to be relatively close to the infrared (IR)
brane.  When the light generations are localized near the ultraviolet
(UV) brane, this can lead to large corrections to the $Zb_{L}\bar{b}_{L}$
coupling, which also impose important restrictions.  A common
solution invokes a custodial symmetry~\cite{Agashe:2003zs}, and a
benchmark model of anarchic warped extra dimensions was proposed in
Ref.~\cite{Agashe:2006hk} (see~\cite{Carena:2003fx} for alternatives).
In that model, the $S$ parameter --together with the one-loop fermion
contributions to $T$ and the $Z b_L \bar{b}_L$
coupling~\cite{Carena:2006bn}, which are calculable in such models--
imposes the strongest constraint: the new vector bosons should be
heavier than $3-4$ TeV.~\footnote{If one gives up on the anarchy
assumption, the constraints can be significantly
relaxed~\cite{Delaunay:2010dw}.} The couplings of the quarks to these
Kaluza-Klein (KK) vector bosons in the previous benchmark model were
taken to be
\begin{equation}
g_{t_R} \sim 5 g_{SM}~, 
\quad\quad
g_{t_L},g_{b_L} \sim  g_{SM}~, 
\quad\quad
g_{q} \sim -g_{SM}/5~, 
\label{couplings:traditional}
\end{equation}
where $q$ represents any of the light Standard Model (SM) fermions and
$g_{SM}$ represents a typical SM couplings.  The conclusion of LHC
reach studies is that the gauge KK resonances are difficult to see at
the LHC, mainly due to their large width and reduced production cross
section, the latter resulting from the small coupling to valence
quarks~\cite{Agashe:2006hk,Fitzpatrick:2007qr}.  Luckily, this
benchmark model comes with another characteristic signature: light
vector-like fermions (fermion custodians)~\cite{Cacciapaglia:2006gp},
that are much easier to see at the LHC~\cite{Contino:2008hi} (see
\cite{Santiago:2011dn} for an overview).

It has been recently suggested~\cite{Cabrer:2010si} that a departure
from pure AdS$_5$ near the infrared brane can substantially decrease
the $T$ parameter (and, although by a smaller amount, also the $S$
parameter, as was previously observed
in~\cite{Falkowski:2008fz,Batell:2008me}).  The authors
of~\cite{Cabrer:2010si} proposed a model in which the reduction of the
$T$ and $S$ parameters is so effective that new vector bosons with
masses below $1$ TeV are compatible with the EWPT \textit{without the
custodial symmetry}.  The analysis in~\cite{Cabrer:2010si} assumes,
however, that all fermions are localized on the UV brane.  This is an
excellent approximation for the light generations but not for the
third one.

We extend the previous study by incorporating bulk fermions and
describing their effects on EWPT once the top and bottom masses are
reproduced.  This is important for a number of reasons: first, without
the custodial symmetry the $Z b_L \bar{b}_L$ coupling is not
protected~\cite{Agashe:2003zs} and can impose significant constraints
on the model; second, the loop-level contributions to the EW
observables are expected to be strongly dependent on the localization
of the third family of quarks, and can also impose significant
constraints; finally, the collider implications of the model depend
crucially on the localization of the third generation quarks.
Nevertheless, we find that the signatures associated with the region
of parameter space favored by the EW precision constraints are still
characterized by Eq.~(\ref{couplings:traditional}), although lighter
resonances than in the AdS$_{5}$ background may be allowed.

The outline of the paper is as follows.  We introduce the model in
Section~\ref{model} and discuss bulk fermions in
Section~\ref{fermions}.  The EW constraints of the model are
investigated in Section~\ref{fit} and its collider implications in
Section~\ref{collider}.  We conclude in Section~\ref{conclusions}.

\section{A Warped Model without Custodial Symmetry}
\label{model}

The model under consideration represents a departure from AdS$_5$ in
the infrared.  We simply summarize in this section the main results
for the gravitational background and the bosonic field content, while
referring the reader to~\cite{Cabrer:2010si} for full details.  We
will then introduce bulk fermions in the model and discuss their
effects on the EWPT. The gravitational background is given by the
metric
\begin{eqnarray}
ds^2 &=& e^{-2A(y)} \eta_{\mu\nu} dx^\mu dx^\nu - dy^2~,
\label{background}
\end{eqnarray}
with warp factor
\begin{eqnarray}
A(y) &=& k y - \frac{1}{\nu^2}\log \left(1-\frac{y}{y_s}\right)~.
\end{eqnarray}
The extra dimension is bounded by two branes: the UV brane, localized
at $y=0$, and the IR brane, localized at $y=y_1$.  The warp factor
would vanish at $y_s$ but we will always consider parameters such that
this metric singularity remains hidden behind the IR brane ($y_s >
y_1$).

The gravitational parameters are therefore $\nu$, $y_1$, $y_s$ and $k$
(the curvature scale at the UV brane, of the order of the Planck mass)
which is assumed to set the scale of all dimensionful 5D quantities.
We will trade $y_s$ for the value of the curvature radius at the IR
brane, given in units of $k$ by
\begin{eqnarray}
k L_1 &=& 
\frac{\nu^2 k(y_s-y_1)}{\sqrt{1-2\nu^2/5+2\nu^2 k(y_s-y_1)
+\nu^4 k^2(y_s-y_1)^2}}~.
\end{eqnarray}
Requiring perturbativity of the gravitational expansion bounds its
value by $kL_1 \gtrsim 0.2$~\cite{Cabrer:2010si}.  The position of the
IR brane, $y_1$, can be fixed by requiring the gravitational
background to generate the $M_P/\mathrm{TeV}$ hierarchy.  We will
simply set $A(y_1)=35$, which determines $y_{1}$.

\begin{figure}[t]
\begin{center}
{
\includegraphics[width=0.55\textwidth,clip=true]{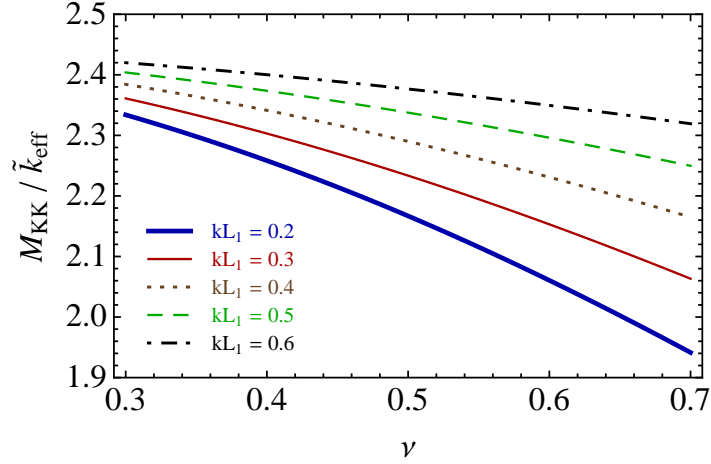}
} 
\caption{Mass of the first gauge KK mode in units of the effective
IR scale $\tilde{k}_{\rm eff}$, defined in Eq.~(\ref{keff}), as a
function of $\nu$ and for different values of $kL_1$.}
 \label{Mkkbyktilde:gauge}
\end{center}
\end{figure}

The bosonic content of the model consists of the SM gauge fields, with
Neumann boundary conditions (BC) on both branes, and an electroweak
scalar doublet, the Higgs.  Other scalars involved in the
gravitational background and the stabilization of the interbrane
distance~\cite{Cabrer:2010si} are irrelevant for this discussion,
although their phenomenology, in particular that of the radion, can be
interesting.  The bulk gauge bosons can be expanded in KK
modes (we focus on the $\mu$ component here, which is the relevant one
for EWPT and collider implications):
\begin{eqnarray}
A_\mu(x,y) &=& \frac{1}{\sqrt{y_1}}\sum_n f_n^A(y) A_\mu^{(n)}(x)~,
\end{eqnarray}
where the profiles satisfy 
\begin{eqnarray}
\left[\partial_y e^{-2A} \partial_y + m_n^2 \right] f_n^A &=& 0~,
\hspace{1cm}
\partial_y f_n^A\big|_{y=0,y_1} ~= ~0~,
\end{eqnarray}
(we treat EWSB perturbatively), and are normalized according to
\begin{eqnarray}
\frac{1}{y_{1}} \int_0^{y_1} \! dy \, f_n^{A}f_m^{A} &=& \delta_{nm}~.
\end{eqnarray} 
The boundary conditions fix the value of the KK masses. They are of
the order of the effective IR scale
\begin{eqnarray}
\tilde{k}_{\rm eff} &\equiv& A^\prime(y_1) \, e^{-A(y_{1})}~,
\label{keff}
\end{eqnarray}
i.e.~of order the warped down curvature at the IR brane.  We show in
Fig.~\ref{Mkkbyktilde:gauge} the mass of the first gauge KK mode, in
units of $\tilde{k}_{\rm eff}$, for different values of $\nu$ and
$kL_1$.
For the Higgs field we can write
\begin{eqnarray}
H(x,y) &=& \frac{1}{\sqrt{2}} \, e^{i\chi(x,y) }
\begin{pmatrix}
0 \\ h(y)+\xi(x,y)\end{pmatrix}~,
\end{eqnarray}
where the Higgs KK expansion reads
\begin{eqnarray}
\xi(x,y) &=& \frac{1}{\sqrt{y_1}} \sum_n f_n^\xi(y) \xi^{(n)}(x)~,
\end{eqnarray}
and the wavefunctions are normalized to
\begin{eqnarray}
\frac{1}{y_{1}} \int_0^{y_1} \! dy \, e^{-2A} f_n^{\xi}f_m^{\xi} &=& \delta_{nm}~.
\label{normHiggs}
\end{eqnarray} 
For a light Higgs we can assume that the Higgs vacuum expectation
value (vev) is carried by the
zero-mode, which has a profile 
\begin{eqnarray}
f_0^\xi(y) &\approx&  N_h \, e^{a k y},
\end{eqnarray}
with $N_h$ fixed by the normalization condition~(\ref{normHiggs}).
Following~\cite{Cabrer:2010si} we trade $a$ for 
\begin{eqnarray}
\delta &\equiv& \left|e^{-2(a-2)k y_s} k y_s [-2(a-2)ky_s]^{-1+\frac{4}{\nu^2}}
\, \Gamma \! \left(
1-\frac{4}{\nu^2},-2(a-2)k(y_s-y_1)\right) \right|~,
\end{eqnarray}
which is a measure of how much fine-tuning in the 5D parameters we
need to impose in order to preserve the Randall-Sundrum solution to
the hierarchy problem~\cite{Randall:1999ee}.

In summary, we have two free parameters for the gravitational
background, $\nu$ and $kL_1$; and one for the Higgs background,
$\delta$.  However the EWPT are quite insensitive to the
latter so we will simply fix $\delta=0.1$ in the following.

\section{Bulk Fermions}
\label{fermions}

We introduce now bulk fermions in the model.  The quadratic part
of the action for bulk fermions reads
\begin{eqnarray}
S &=& \int \! d^4x \, dy \, e^{-4A} \, \overline{\Psi}\Big[e^A i \cancel{\partial}
  +(\partial_5 -2 A^\prime)\gamma^5 - M \Big]\Psi~,
\end{eqnarray}
where $M$ is a 5D Dirac mass allowed by the symmetries.  Following the
standard procedure we expand the bulk fermions in KK modes
\begin{eqnarray}
\Psi_{L,R}(x,y) &=& \frac{1}{\sqrt{y_1}} \sum_n f_n^{L,R}(y) \psi_{L,R}^{(n)}(x)~,
\end{eqnarray}
where we have defined the 4D chiralities $\Psi_{L,R}= \frac{1}{2}
(1\mp \gamma^5) \Psi$, and the KK profiles satisfy the following
system of equations:
\begin{eqnarray}
\left[\partial_y - \left(2 A^\prime+M\right)\right]f_n^R &=&- e^A m_n f_n^L~,
\nonumber \\
\left[\partial_y - \left(2 A^\prime-M\right)\right]f_n^L &=& e^A m_n f_n^R~,
\end{eqnarray}
with the ortho-normality conditions given by 
\begin{eqnarray}
\frac{1}{y_{1}} \int_0^{y_1} \! dy \, e^{-3A} f_n^L f_m^L &=&
\frac{1}{y_{1}} \int_0^{y_1} \! dy \, e^{-3A} f_n^R f_m^R ~=~ \delta_{nm}~.
\end{eqnarray}
Bulk fermions with a left-handed (LH) zero-mode are obtained by
imposing Dirichlet boundary conditions for their right-handed (RH)
chirality on both branes.  Such bulk fermions will be denoted as
``$[++]$ fermions''.  The boundary conditions for the opposite, LH,
chirality are fixed by the equations of motion.  Similarly, bulk
fermions with a RH zero-mode are obtained by imposing Dirichlet
boundary conditions on their LH chirality, and will be denoted as
``$[--]$ fermions''.  The LH (RH) massless profile of a $[++]$
($[--]$) bulk fermion is given by
\begin{eqnarray}
f_0^{L,R}(y) &=& N_0^{L,R} \, e^{2A\mp \int_0^y \! dy' M(y')}~,
\end{eqnarray}
where $N_0^{L,R}$ is fixed by the normalization
condition.  In the following we will consider a constant bulk mass
$M_i=c_i k$, where $i$ denotes the fermion type.~\footnote{In the
limit in which $y_s \to y_1$ there could be problems of strong
coupling similar to the ones present in soft-wall
models~\cite{Batell:2008me}, and a $y-$dependent mass term might be
necessary~\cite{MertAybat:2009mk} (see~\cite{Gherghetta:2009qs} for
other realizations of flavor in soft-wall models).}

\begin{figure}[t]
\centerline
{
\includegraphics[width=0.55\textwidth,clip=true]{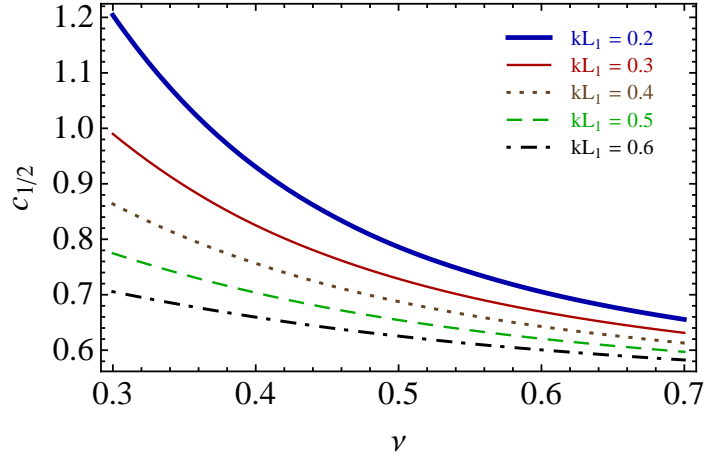}
} 
\caption{ Value of the localization parameter, $c_{1/2}$, that makes
a LH fermion zero-mode mostly delocalized, as a function of $\nu$ and 
for different values of $kL_1$.}
\label{c0}
\end{figure}

In the present metric background there are no exactly flat fermion
solutions, unlike for a pure AdS$_{5}$ background, $A_{\rm RS}(y) =
ky$, with a constant 5D Dirac mass defined by $c=1/2$.  However, the
$c$ parameter does control the localization of the fermion zero-modes,
either towards the IR or UV brane.  In particular, there is always a
background-dependent value, $c_{1/2}$, such that for $c < c_{1/2}$,
the (LH) fermion zero-mode is mostly IR localized, while for $c >
c_{1/2}$ it is mostly UV localized.  This value plays a role analogous
to $c = 1/2$ in the pure AdS$_{5}$ background.  In Fig.~\ref{c0} we
show the value of $c_{1/2}$, defined such that $e^{-\frac{3}{2}A(0)}
f^{L}_{0}(0) = e^{-\frac{3}{2}A(y_{1})} f^{L}_{0}(y_{1})$ (see
Eq.~(\ref{omegas}) and footnote~\ref{physical}), as a function of the
input parameters $\nu$ and $kL_1$ (see also
Fig.~\ref{localization:fig}).

\begin{figure}[t]
\centerline
{
\includegraphics[width=0.49\textwidth,clip=true]{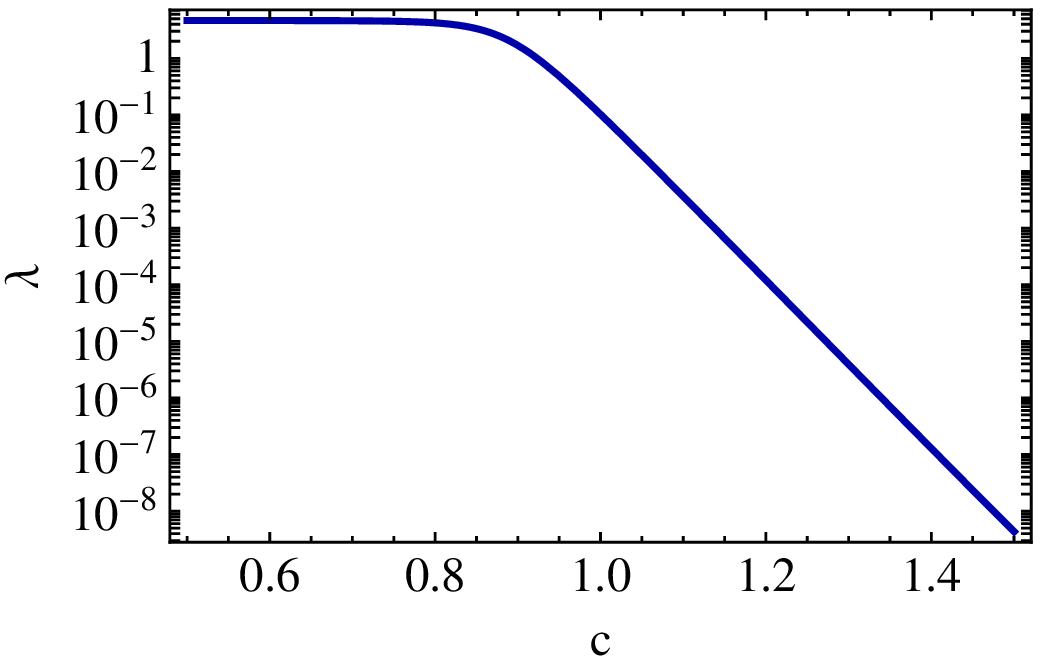}
\hspace{0.02 \textwidth}
\includegraphics[width=0.455\textwidth,clip=true]{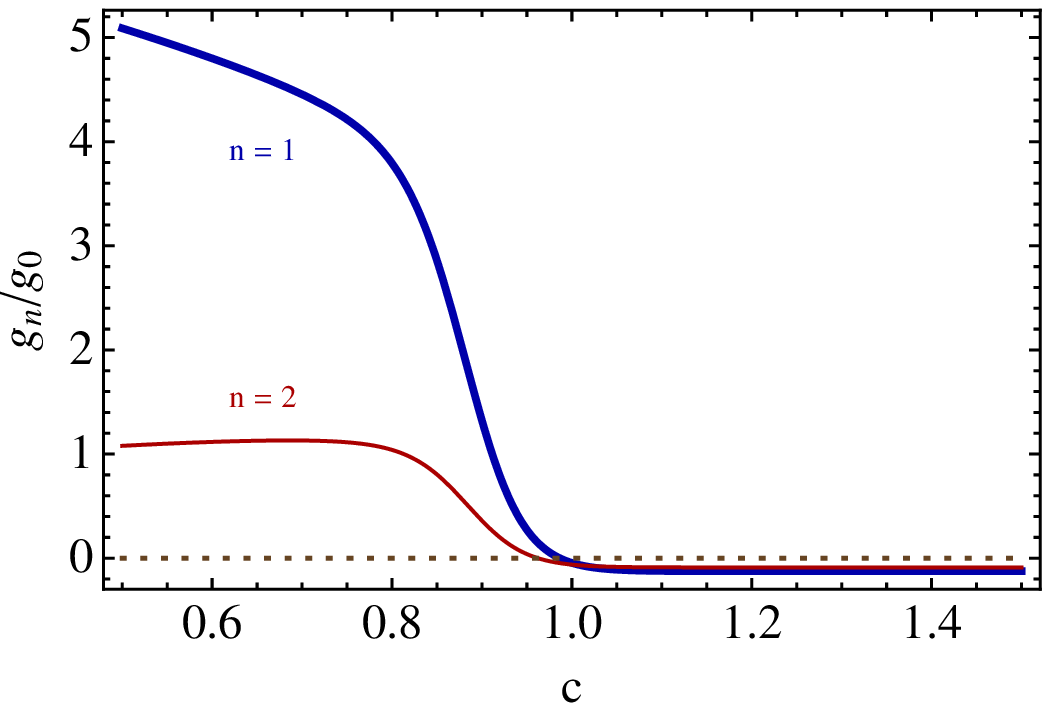}
} 
\caption{Left panel: Yukawa coupling, $\lambda \equiv
\lambda^{LR}_{00}$, as a function of the bulk mass for $c_L=-c_R=c$
and a 5D Yukawa coupling that saturates the maximum value given in
Eq.~(\ref{Y5NDA}).  Right panel: coupling of a LH fermion zero-mode to
the first two gauge boson KK modes in units of $g_{0} \equiv
g_5/\sqrt{y_1}$.  In both cases we have taken $\nu=0.4$ and
$kL_1=0.2$, which lead to maximally delocalized fermions for $c
\approx 0.93$.}
\label{yukandcoupvsc:fig}
\end{figure}

Let us now discuss the couplings of the fermion zero-modes to the
Higgs and gauge boson KK modes in this model.  The gauge couplings are
given by
\begin{eqnarray}
\mathcal{L}&\supset&g_5 \int \! dy \, e^{-3A} \, \overline{\Psi} \, \cancel{\!A} \Psi
~=~ \sum_{mnr} \frac{g_5}{\sqrt{y_1}} \int  \! dy \, e^{-3A} \, \frac{f_n^L f_m^L f_r^A}{y_1} \,\, 
\overline{\psi}_L^{(n)} \cancel{\!A}^{(r)} \psi_L^{(m)} + (L \to R)
\nonumber \\
&\equiv& \sum_{mnr} g^L_{nmr} \overline{\psi}_L^{(n)} \cancel{A}^{(r)}
\psi_L^{(m)} + (L \to R)
~, 
\end{eqnarray}
whereas the Yukawa couplings can be computed from
\begin{eqnarray}
\mathcal{L} &\supset& Y_5 \int \! dy \, e^{-4A} \, \bar{Q}_L H \, U_R + (L
\leftrightarrow R) 
\\
&=& \sum_{mn} \frac{Y_5}{\sqrt{y_1}} \int  \! dy \, e^{-4A}  \,  \frac{f_L^{Q(n)} f_\xi^{(0)} 
f_R^{U(m)}}{y_1} \,\, \bar{q}_{L}^{(n)} h \, u_R^{(m)} +
(L\leftrightarrow R)+ \cdots 
\nonumber 
~\equiv~ \lambda^{LR}_{mn} \, \bar{q}_{L}^{(n)} h \, u_R^{(m)} + \cdots~,
\end{eqnarray}
where we have defined $h \equiv \xi^{(0)}(x)$.  The 5D gauge coupling
can be fixed by matching the coupling of the gauge zero-mode to the
observed 4D coupling.  Assuming a tree-level matching this gives
\begin{eqnarray}
g_5 &=& \sqrt{y_1} \, g_4~.
\end{eqnarray}
For the 5D Yukawa coupling, Ref.~\cite{Agashe:2008uz} finds, based on
NDA~\cite{Chacko:1999hg}, that its maximum value is given by
\begin{eqnarray}
Y_5 &\leq& Y_5^{\mathrm{max}} ~\approx~ \frac{4\pi}{\sqrt{3 k}}~,
\label{Y5NDA}
\end{eqnarray}
which corresponds to strong coupling at a scale of the third KK level.
We plot in the left panel of Fig.~\ref{yukandcoupvsc:fig} the Yukawa
coupling between the zero-modes of a $[++]$ fermion $Q$, and a $[--]$
fermion $T$, assuming $Y_5=Y_5^{\mathrm{max}}$, as a function of a
common localization parameter $c=c_Q=-c_T$.  In the right panel we
show the coupling of a LH fermion zero-mode to the first two gauge
boson KK modes divided by the coupling to the gauge boson zero-mode
(assuming tree level matching) as a function of $c$.  We have chosen
$\nu=0.4$, and $k L_1=0.2$ in both plots.  We see the change of
behavior around $c_{1/2} \approx 0.93$, which corresponds to the most
delocalized zero-mode fermion profile (see Fig.~\ref{c0}).

For $c \gtrsim c_{1/2}$, we see that the Yukawa coupling becomes
exponentially suppressed.  Furthermore, in the same region the
coupling to the gauge boson KK modes becomes almost universal (i.e.
independent of $c$).  Thus, as for the pure AdS$_{5}$ case, the
assumption that the light fermions are exactly localized on the UV
brane is an excellent approximation (as far as EWPT are concerned) in
these models.
%

\section{Electroweak Constraints}
\label{fit}

If flavor is explained by means of localization in the extra
dimension, the effects of new physics in models with warped extra
dimensions are almost universal and can therefore be encoded in the
oblique parameters of~\cite{Barbieri:2004qk}, with the most important
exception being the couplings of the bottom quark.  This can be most
easily seen with holographic methods.  However, the collider
implications of these models are easier to understand if we discuss
how the corrections are generated in the physical basis (i.e.~the
KK-basis).  A general discussion of EWPT in models with warped extra
dimensions and the equivalence of different methods to compute them is
given in~\cite{Davoudiasl:2009cd}.  Here we follow the equivalent
notation in~\cite{Cabrer:2010si} to make the comparison easier.
Before presenting the detailed EW precision analysis, we comment on
the calculability of such effects in the present class of models.

\subsection{Calculability} 
\label{Calculability}

\begin{figure}[t]
\centerline
{
\includegraphics[width=0.65\textwidth]{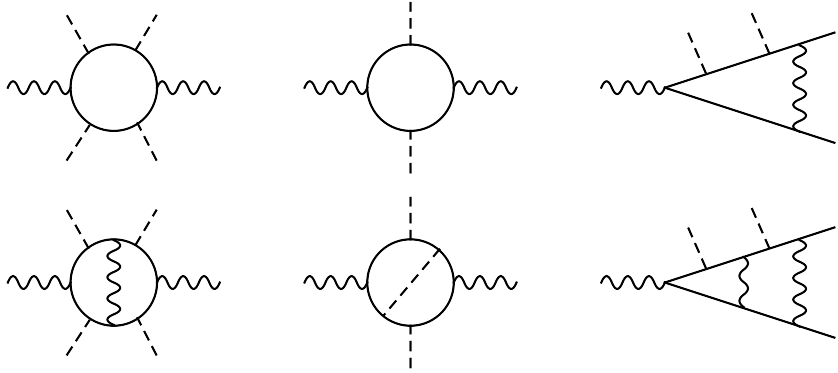}
}
\caption{Examples of radiative corrections to the oblique
parameters, as well as to the non-oblique vertex corrections.  The
most important contributions arise from the top KK-tower in the loop.
Upper row, left to right: one-loop contributions to $T$, $S$ and
$\delta g_{b_{L}}$, respectively.  Lower row: examples of the
corresponding two-loop contributions.
\label{fig:LoopCorrections}}
\end{figure}

As already emphasized, the present class of models does not have a
custodial symmetry to protect the Peskin-Takeuchi $T$ parameter, nor
the corrections to certain gauge-fermion couplings.  As a result, it
is possible to write a term in the bulk that violates the custodial
symmetry, ${\cal L}_{5} \supset (\kappa/\Lambda^3) \, |H^{\dagger}
D_{M} H|^{2}$, where we wrote the coefficient in units of the 5D
cutoff, $\Lambda$.  The dimensionless factor, $\kappa$, is UV
sensitive.  However, when the Higgs propagates in the bulk, the 1-loop
contributions are finite by power-counting.  To see this, it is
simplest to use a normalization where the gauge bosons, $W_{M}$, have
mass dimension 1 [we write the kinetic term as $(-1/4g_{5}^{2}) \,
W^a_{MN} W^{a\,MN}$].  Then, the contributing diagrams are simply
proportional to $g_{5}^{4}$ or $y_{5}^4$, where both the 5D (top)
Yukawa coupling, $y_{5}$, and the 5D gauge coupling have mass
dimension $-1/2$.  (We show in the upper-left corner of
Fig.~\ref{fig:LoopCorrections} an example diagram with a fermion
loop.)  It follows that the piece proportional to $\eta_{\mu\nu}$ in
the loop integral has mass dimension $-1$, and therefore it is IR
dominated.  Subtracting the zero-mode contribution, the remainder is
controlled by the scale $\tilde{k}_{\rm eff}$.  At two-loop order, an
example of which is shown in the lower-left corner of
Fig.~\ref{fig:LoopCorrections}, the diagram is
\textit{logarithmically} divergent by power counting.

A similar point has been made in connection to the $S$-parameter when
the Higgs is taken as a bulk field.\footnote{K. Agashe, private
communication.  See also~\cite{Delaunay:2010dw}.} The middle diagram
in the upper row of Fig.~\ref{fig:LoopCorrections} is finite, while at
2-loop order a \textit{logarithmic} UV sensitivity is encountered
(e.g.~in the middle diagram of the lower row of
Fig.~\ref{fig:LoopCorrections}), corresponding to the bulk operator
$H^{\dagger} \sigma^{a} H \, W^{a}_{MN} B^{MN}$.  Similar remarks
apply to ``vertex corrections'' of the form $(H^{\dagger} D_{M} H) (
\overline{\Psi} \, \Gamma^M \Psi)$ or $(H^{\dagger} \sigma^{a} D_{M}
H) ( \overline{\Psi} \, \Gamma^M \sigma^{a} \Psi)$.  Examples of 1-
and 2-loop contributions to these vertex corrections are shown in the
right column of Fig.~\ref{fig:LoopCorrections}.  We also note that
operators localized on the branes (corresponding either to $T$, $S$ or
$\delta g_{b_{L}}$) can be induced only at three (and
higher-order) loop order.

We conclude that just allowing the Higgs to propagate in the bulk can
make the oblique parameters effectively calculable: the incalculable
pieces associated with the 5D local operators above are suppressed
compared to the finite, one-loop contribution roughly by $M_{\rm
KK}/\Lambda$.  In the particular models studied here, the localization
properties of the Higgs, to be discussed in the next subsection, imply
that the couplings of the Higgs to the KK fermion or gauge states are
suppressed compared to the situation in an AdS$_{5}$ background.  In
fact, for the favored region of parameter space such couplings are
well in the perturbative regime (thus resembling more closely the case
of flat rather than typical warped extra-dimensions).  However,
although higher-order contributions due to Yukawa or weak gauge
couplings can be expected to be further suppressed, there remains some
uncertainty associated with higher order QCD contributions.  As we
will see in detail below, one of the effects of the departure from
AdS$_5$ of the gravitational background is to push the KK modes closer
to the IR brane.  This effect increases the coupling among KK modes,
thus reducing the scale of strong coupling in the QCD sector.  This is
reminiscent of the position dependent cut-off in soft-wall
models~\cite{Falkowski:2008fz,Batell:2008me}, and affects the EW
observables at two-loop and higher order.\footnote{A smaller hierarchy
in the spirit of the Little RS model~\cite{Davoudiasl:2008hx} would
reduce the tension with the low QCD cut-off and increase the coupling
of the KK-gluon to light quarks, thus having an important impact on
collider phenomenology.} Nevertheless, we find that the finite
one-loop contributions to the EW observables --especially to $T$-- can
be significant, and therefore one should not neglect such radiative
effects.  To get a concrete idea about their impact on the EW fit and
the resulting bounds on the KK scale, we will assume that the QCD
strong coupling scale is high enough to make the 5D description a
reasonable approximation.  Furthermore we will assume that 2-loop and
higher-order QCD effects do not dramatically change the one-loop
results.  One should, however, keep in mind that QCD effects may not
be negligible.

\subsection{The Oblique Corrections at Tree Level} 
\label{TreeLevel}

We perform now a detailed analysis of the EW precision constraints in
the class of models without custodial symmetries under discussion.  In
this subsection we focus on the \textit{oblique} analysis at
\textit{tree level} (which was already performed
in~\cite{Cabrer:2010si}), and in the next one we take the most
important non-oblique contribution into account, i.e.~the correction
to the $Zb_{L}\bar{b}_{L}$ coupling, as well as the one-loop
contributions.  This will allow us to better understand the impact of
the various effects.  Let us start with the gauge KK-modes.  The
effects of new physics can be classified in three types: corrections
to the gauge boson self-energies, to fermion-gauge couplings and to
four-fermion interactions.  Each of these effects is characterized by
coefficients denoted by $\hat{\alpha}$, $\hat{\beta}$ and
$\hat{\gamma}$, respectively.  For the present case, only gauge bosons
obeying Neumann boundary conditions on both branes are relevant, in
which case one has~\cite{Cabrer:2010si}
\begin{eqnarray}
\hat{\alpha}&=&
\int_{0}^{y_1} \! dy \, e^{2A(y)}\left(\Omega_h(y)
-\frac{y}{y_1}\right)^2~,
\nonumber \\
\hat{\beta}&=&
\int_{0}^{y_1} \! dy \, e^{2A(y)}
\left(\Omega_h(y)-\frac{y}{y_1}\right)
\left(\Omega_f(y)-\frac{y}{y_1}\right)~,
\\
\hat{\gamma}&=&
\int_{0}^{y_1} \! dy \, e^{2A(y)}
\left(\Omega_f(y)-\frac{y}{y_1}\right)^2~.
\nonumber
\label{alphabetagamma:eq}
\end{eqnarray}
The function $\Omega(y)$ is defined by~\footnote{\label{physical}The
$\omega_{i}(y)$ are nothing but the (square of the) ``physical
wavefunction'' profiles, i.e.~the profiles with warp factors taken
out, as dictated by dimensional analysis (which just redshift all mass
scales appropriately).}
\begin{eqnarray}
\Omega_i(y) &\equiv& \frac{1}{y_1} \int_0^{y} \! d\tilde{y} \,
\omega_i(\tilde{y}),
\qquad
\omega_i(y) ~\equiv~ \left \{
\begin{array}{l}
e^{-2A(y)} f^2_{i,0}(y)~, \quad \mathrm{scalars} \\
e^{-3A(y)} f^2_{i,0}(y)~, \quad \mathrm{fermions} 
\end{array}
\right.~,
\label{omegas}
\end{eqnarray}
where $f_{i,0}(y)$ is the wave function for the scalar or fermion
zero-modes (see previous sections).  With our normalization we have
$\Omega_i(y_1)=1$.
\begin{figure}[t]
\centerline
{
\includegraphics[width=0.23\textwidth,clip=true]{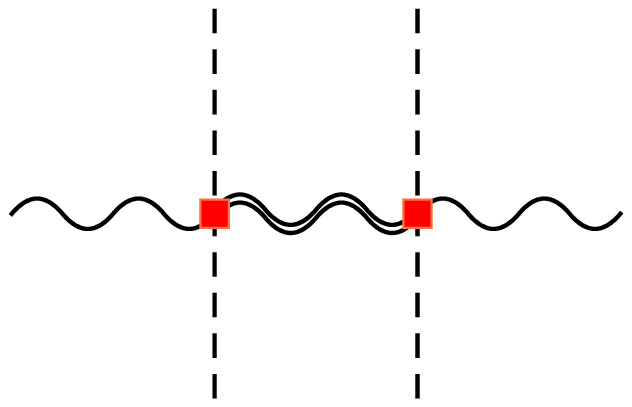}
\hspace{0.05 \textwidth}
\includegraphics[width=0.23\textwidth,clip=true]{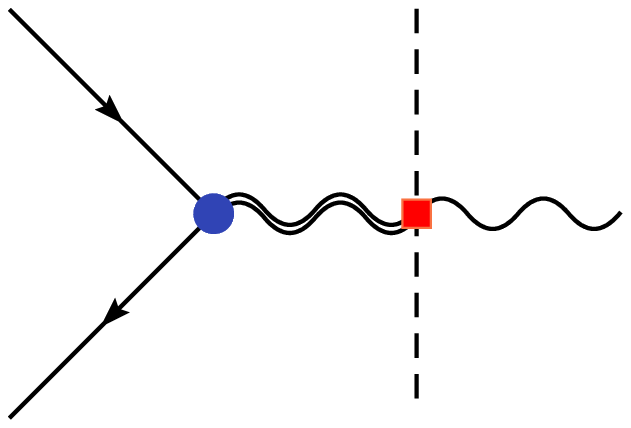}
\hspace{0.05 \textwidth}
\includegraphics[width=0.23\textwidth,clip=true]{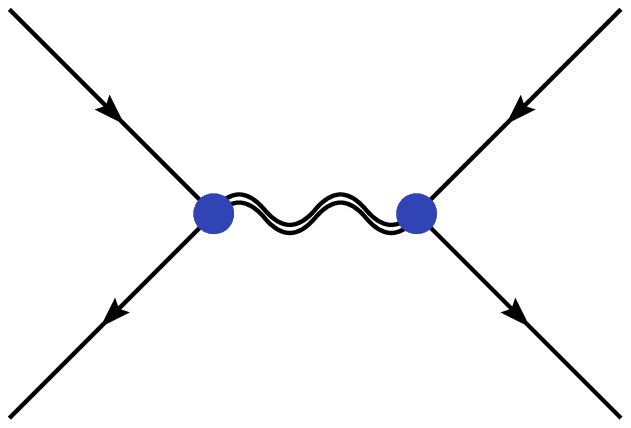}
} 
\caption{Contribution to the coefficients $\hat{\alpha}$,
$\hat{\beta}$ and $\hat{\gamma}$ of Eq.  (\ref{alphabetagamma:eq}).
The double line represents the tower of massive gauge boson KK modes,
the red squares the mixing between the gauge KK and zero-modes, and
the blue dot the coupling of the light fermions to the gauge KK modes.
 \label{alphabetagamma:fig}}
\end{figure}
These coefficients are diagrammatically shown in
Fig.~\ref{alphabetagamma:fig}.  The wavy double line represents the
tower of \textit{massive} gauge boson KK modes, resummed in
Eqs.~(\ref{alphabetagamma:eq}).  The blue dot represents the coupling
of the SM fermions to the new vector bosons, and the red square the
mixing between SM and heavy gauge bosons through Higgs insertions.
The oblique parameters due to the KK physics can be written in terms
of these coefficients as
\begin{eqnarray}
\hat{T}&=&\frac{g^{\prime\, 2}v^2}{2}(\hat{\alpha}-2\hat{\beta}+
\hat{\gamma})~, \\
\hat{S}&=& g^2 v^2(-\hat{\beta}+\hat{\gamma})~, \\
W&=& Y=\frac{g^2v^2}{2} \hat{\gamma}~,
\end{eqnarray}
where $g$ and $g^\prime$ are the $SU(2)_L$ and $U(1)_Y$ couplings,
respectively, and $v=174$ GeV is the Higgs vev.  These are related to
the Peskin-Takeuchi $S$ and $T$ parameters~\cite{Peskin:1991sw} by
$\alpha S = 4 s^{2}_{W} \hat{S}$ and $\alpha T = \hat{T}$.  We also
include a Higgs contribution to $S$ and $T$, given
by~\cite{Veltman:1976rt}
\begin{eqnarray}
\Delta S &=& \frac{1}{2 \pi} \left[ g(m_{h}^2/m^2_{Z}) - g(m_{\rm ref}^2/m^2_{Z}) \right]~,
\\[0.5em]
\Delta T &=& - \frac{3}{16 \pi c_{W}^2} \left[ f(m_{h}^2/m^2_{Z}) - f(m_{\rm ref}^2/m^2_{Z}) \right]~,
\end{eqnarray}
where
\begin{eqnarray}
g(y) &=& \int_{0}^1 \! dx \, x (5 x - 3) \ln(1 - x + y x)~,
\\[0.5em]
f(y) &=& y \, \frac{\ln c_{W}^2 - \ln y}{c_{W}^2 - y} + \frac{\ln y}{c_{W}^2 (1 - y)}~,
\end{eqnarray}
and we take $m_{\rm ref} = 115~{\rm GeV}$ in the fit.

\begin{figure}[t]
\centerline
{
\includegraphics[width=0.55\textwidth,clip=true]{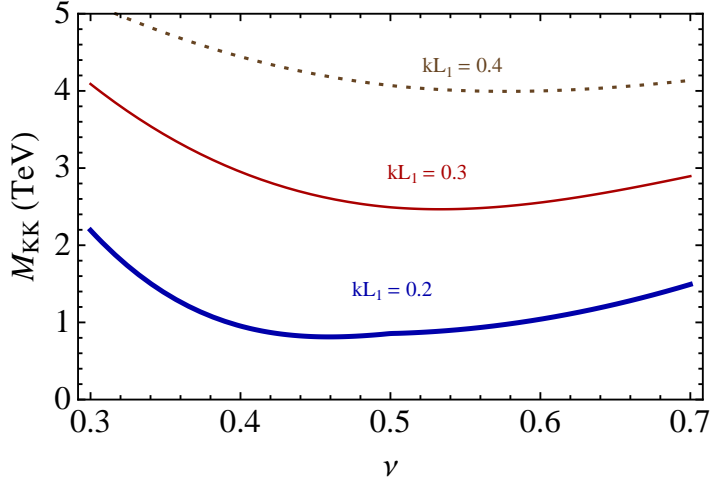}
} 
\caption{$95 \%$ CL lower bound on the mass of the first KK gauge
boson, obtained from a \textit{tree-level oblique} analysis.  We
assume UV localized fermions and different values of the input
parameters $\nu$ and $kL_1$.  See text for details on the fit
procedure.  \label{mKKbosonic:fig}}
\end{figure}
We show in Fig.~\ref{mKKbosonic:fig} the result of a fit to the
oblique parameters when all fermions are assumed to live on the UV
brane for different values of the input parameters $\nu$ and $kL_1$.
We have used an updated version of the code in~\cite{Han:2004az},
obtaining the bounds as follows.  For fixed values of $\nu$ and $kL_1$
(i.e.~for a fixed metric background), we compute the minimum of the
$\chi^2$, including only the $Z$-pole observables,\footnote{These are
the 26 observables associated with the W mass (two measurements), the
Z-line shape and lepton FB asymmetries (8), heavy flavor (6),
effective $\sin^{2}\theta_{W}$ (2), and 8 leptonic/strange quark
polarization asymmetries.  We do not fit the SM parameters, but only
those associated with the new KK physics.} as a function of the Higgs
mass (imposing the direct LEP bound of $\approx 114~{\rm GeV}$) and
the value of the IR scale $\tilde{k}_{\rm eff}$.  We then fix the
Higgs mass to its value at the minimum and compute the bound on
$\tilde{k}_{\rm eff}$ by requiring $\Delta \chi^2=3.84$ (95\% CL for
one degree of freedom).  The resulting value of the first gauge boson
KK mode mass (see Fig.~\ref{Mkkbyktilde:gauge}) is plotted as a
function of $\nu$ for different values of $kL_1$.
We see that masses as light as $\sim 1$ TeV are allowed for $kL_1=0.2$
and $\nu\sim 0.45$.  This plot reproduces the results presented
in~\cite{Cabrer:2010si} up to small differences ($\lesssim 100$ GeV),
due to the different fit procedure.

\begin{figure}[t]
\centerline
{
\includegraphics[width=0.65\textwidth,clip=true]{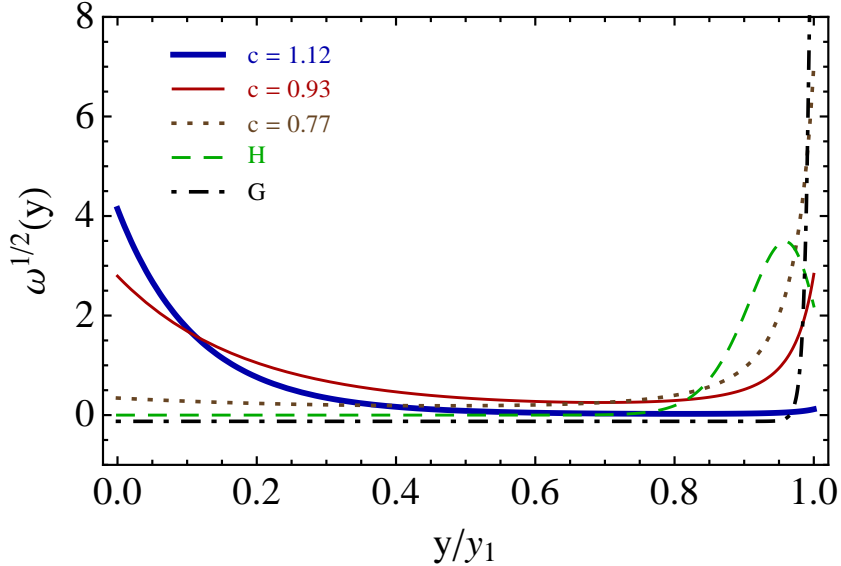}
} 
\caption{Fermion zero-mode, Higgs and first gauge KK mode profiles,
$\sqrt{\omega}(y)$ for $\nu=0.4$ and $kL_1=0.2$.}
\label{localization:fig}
\end{figure}
The reason for the reduced lower bound on the mass of the gauge KK
modes, compared to the pure AdS$_{5}$ background, despite the absence
of custodial symmetry in the model, is easy to understand.  In the
AdS$_{5}$ case, the warp factor forces the gauge boson KK modes to be
localized close to the IR brane, where the Higgs is also localized.
The large overlap then makes the mixing between the gauge boson
zero-mode and massive KK-modes large (the red square in
Fig.~\ref{alphabetagamma:fig}).  The light fermions on the other hand
are localized towards the UV brane and therefore their coupling to the
gauge boson KK modes (the blue dot) is typically small.  This is the
reason for the usual enhancement of the $\hat{T}$ parameter
($\hat{\alpha}$ is proportional to the gauge mixing squared) over the
$\hat{S}$ parameter, and of $\hat{S}$ over $W$ and $Y$.  The departure
from the AdS$_5$ background has several effects that go in the right
direction to improve the EWPT. The first is that the KK gauge bosons
are more strongly localized towards the IR brane whereas the Higgs,
although localized towards the IR brane, reaches a maximum
\textit{before} the IR brane [the ``physical'' Higgs wavefunction is
given by $\sqrt{\omega_{h}(y)} \approx e^{-A(y)} \, e^{a k y}$, and
the suppression near the IR brane arises from the nearby singularity
in $A(y)$].  This reduces, sometimes dramatically, the gauge mixing
through the Higgs vev, and therefore the contribution to the $\hat{T}$
parameter (and somewhat less the $\hat{S}$ parameter).
This effect can be observed in Fig.~\ref{localization:fig}, in which
we show the ``physical'' profiles $\sqrt{\omega}$ of
Eq.~(\ref{omegas}), for the Higgs, the first gauge KK mode, and for
bulk fermion zero-modes with three different values of the bulk
masses.  The maximum of the Higgs profile before the IR brane is
evident, together with the very strong localization of the first gauge
KK mode, leading to a reduced value of the relevant overlap integral.

\begin{figure}[t]
\centerline
{
\includegraphics[width=0.6\textwidth,clip=true]{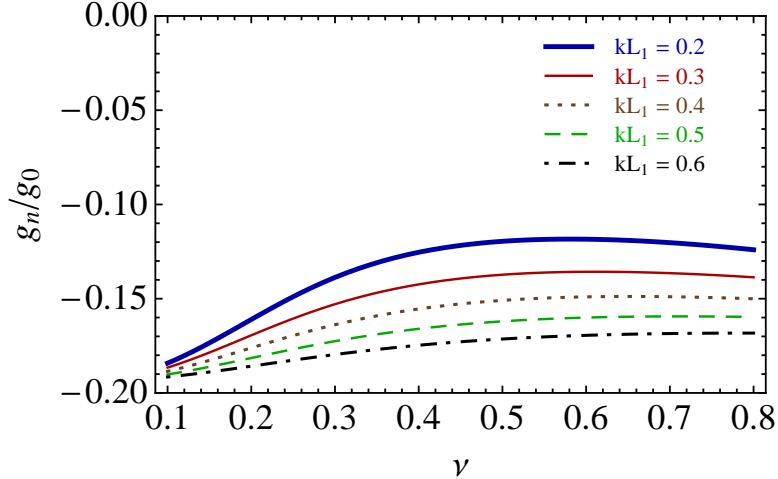}
} 
\caption{Couplings of UV localized fermions to the first-level KK
gauge bosons, in units of the zero-mode gauge coupling.  We show
curves as a function of $\nu$, and for different values of $kL_{1}$.}
\label{glightvsnu:fig}
\end{figure}

The second effect of the warp factor is that the coupling of UV
localized fermions to gauge KK modes is reduced with respect to the
standard AdS$_{5}$ case.  This further reduces the $\hat{S}$ parameter
(but will also have a negative impact on collider searches).  In
Fig.~\ref{glightvsnu:fig} we show the value of the coupling of UV
localized fermions to the first gauge boson KK mode (in units of
$g_{0} = g_5/\sqrt{y_1}$) for different values of $\nu$ and $kL_1$.
Only for small values of $\nu$ and large values of $kL_1$ does one
approach the coupling of the AdS$_{5}$ background.  Note that for the
values preferred by the EWPT (see Fig.~\ref{mKKbosonic:fig}) this
coupling is reduced to almost half the AdS$_{5}$ value.
%

\subsection{The Effects of the Third Generation}
\label{ThirdGeneration}

Let us now consider the effect of the third quark generation.  We
consider three bulk fermion fields with the following quantum numbers
under the $SU(2)_L \times U(1)_Y$ gauge group and BC
\begin{equation}
Q=(2,1/6)\sim [++]~,
\quad
T=(1,2/3)\sim [--]~,
\quad
B=(1,-1/3)\sim [--]~,
\end{equation}
with localization parameters $c_{Q}$, $c_{T}$ and $c_{B}$,
respectively.  For fixed values of these localization parameters we
can choose the 5D Yukawa couplings so that the top and bottom masses
are reproduced.  Due to the upper bound on $Y_5$, the top mass cannot
be generated if the $Q$ and $T$ zero-modes are far from the IR brane.
The LH bottom, which is in the zero-mode of $Q$, can then receive
large corrections to its couplings, from gauge and fermion KK modes.
Anomalous contributions to the $Z b_L \bar{b}_L$ coupling can
therefore impose stringent constraints in warped models without
custodial protection.  The general expression for the
\textit{tree-level} correction to the $Z b_L \bar{b}_L$ coupling
induced by the gauge boson KK modes is given
in~\cite{Davoudiasl:2009cd}.  For our model it reads
\begin{eqnarray}
\delta g_{b_L} &=&
-\frac{g^2v^2}{2c_W^2} \left[\frac{g^2}{2}+\frac{g^{\prime\,2}}{6}\right](\hat{\beta}_Q-\hat{\beta}_{UV})~,
\end{eqnarray}
where the term proportional to $\hat{\beta}_{UV}$ corresponds to the
universal part that is absorbed in the oblique parameters.  To this we
have to add the \textit{tree-level} fermionic contribution, that we
have computed exactly by diagonalizing numerically the KK fermion mass
matrix, and computing the resulting coupling to the $Z$ of the
lightest mass eigenstate (the bottom quark).  We will consider the
1-loop effects separately.

In order to test the dependence of the constraints on the assumptions
on the 5D Yukawa coupling we have taken the following benchmark
scenarios.  For each value of $\nu$ and $kL_1$ and each value of $c_Q$
we fix $c_T$ so that the top mass is reproduced (including the effect
of the mixing with fermion KK modes) assuming that
$Y_5^t=Y_5^{\mathrm{max}}$ (\textbf{scenario 1}).  We note that, due
to the maximum of the Higgs profile, there is a fixed value of $c_T$
for which the overlap is maximal and therefore the 5D Yukawa coupling
is minimal.  Therefore, we also consider the case that $Y_5^t$ is the
minimal one for which it is possible to reproduce the correct top mass
(\textbf{scenario 2}).  Regarding the bottom sector, we consider three
different scenarios by fixing the 5D bottom Yukawa to $Y_5^b=Y_5^t$
(\textbf{scenario \textit{a}}), $Y_5^b=Y_5^t/5$ (\textbf{scenario
\textit{b}}), $Y_5^b=Y_5^t/10$ (\textbf{scenario \textit{c}}).
Scenario $a$ assumes exact anarchy whereas in the other two we allow
for deviations between different Yukawas.

\begin{figure}[t]
\centerline
{
\includegraphics[width=0.5\textwidth,clip=true]{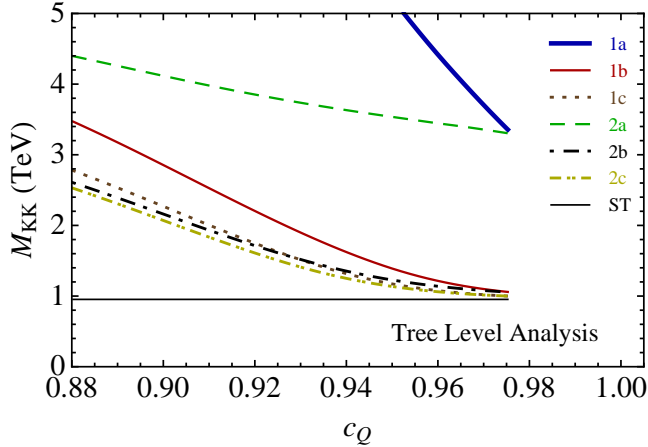}
} 
\caption{Tree-level $95\%$ CL lower bound on the mass of the first
gauge KK mode as a function of the localization of the LH top/bottom
multiplet ($c_Q$) for the six scenarios discussed in the text.  We fix
$\nu=0.4$ and $kL_1=0.2$.  The lines end at the value of $c_Q$ beyond
which the top mass cannot be generated.  For comparison we also show
the result of the fit if the effects of third generation quarks are
neglected (horizontal ``ST'' line).  The maximal delocalization is
obtained for $c \approx 0.93$ (see Fig.~\ref{c0}).
 \label{fitsample:fig}}
\end{figure}
We have studied the tree-level effect of third generation quarks on
EWPT by performing a scan over $\nu$ and $kL_1$.  For each value of
these parameters, we have computed the maximum value of $c_Q$ that
allows to generate the top mass (i.e.~the furthest from the IR brane).
We have then scanned over the values of $c_Q$ smaller than this
maximal value for the six different scenarios ($1a,1b,1c,2a,2b,2c$)
described above.  The fit has been performed as in the case of the
gauge contribution, described in the previous subsection, but
including now the constraint from the $Z b_L \bar{b}_L$ coupling
described above.  Fixing $c_{Q}$ and marginalizing over the Higgs
mass, we find the value of $\tilde{k}_{\rm eff}$ that gives $\Delta
\chi^2=3.84$, corresponding to $95\%$ CL for one degree of freedom.
In all cases the preferred value of the Higgs mass is close to its
current lower limit $m_h\approx 114$ GeV. We show in
Fig.~\ref{fitsample:fig} a sample result for $\nu=0.4$ and $kL_1=0.2$.
We display, based on a tree-level analysis, the $95\%$ CL lower bound
on the mass of the first gauge KK mode as a function of $c_Q$ for all
six scenarios, and also the result of the fit when all fermions are
localized on the UV brane as considered in~\cite{Cabrer:2010si}
(horizontal line denoted by ST in the plot).
It is clear that the bound is very sensitive to the value of the
localization of the LH top/bottom doublet as expected for a
non-custodial model.  The correction is smaller for a larger $c_Q$
(the LH top/bottom further from the IR brane).  The non-trivial result
we find is that the top mass (just barely) allows $c_Q$ to be large
enough to make the corrections to the $Zb_L\bar{b}_L$ coupling
negligible (the lines in the figure end at the point of $c_Q$ beyond
which the top mass cannot be generated).  Unfortunately, loop effects
change this conclusion, as discussed below.  The other property that
is clear from the plot is that exact anarchy (scenarios $1a$ and $2a$)
is extremely constrained by EWPT. The reason is that the Yukawa
couplings in the bottom sector are very large and the mixing with the
bottom KK modes induces very large corrections to the $Z b_L
\bar{b}_L$ coupling.  However, we see that a suppression in the 5D
bottom Yukawa by a factor of 5 is already enough to get the bound on
the scale of new physics reasonably low.  We have also checked that
the $\bar{t}_R \gamma^\mu b_R W_\mu^+$ coupling is in these cases
typically $\lesssim 10^{-4}$ (in units of $g/\sqrt{2}$) and should
therefore cause no trouble with $b\to s \gamma$
constraints~\cite{Cho:1993zb} (a full analysis, including the fermion
KK modes in the calculation, to properly account for this constraint,
is beyond the scope of this paper).

\begin{figure}[t]
\centerline
{
\includegraphics[width=0.53\textwidth,clip=true]{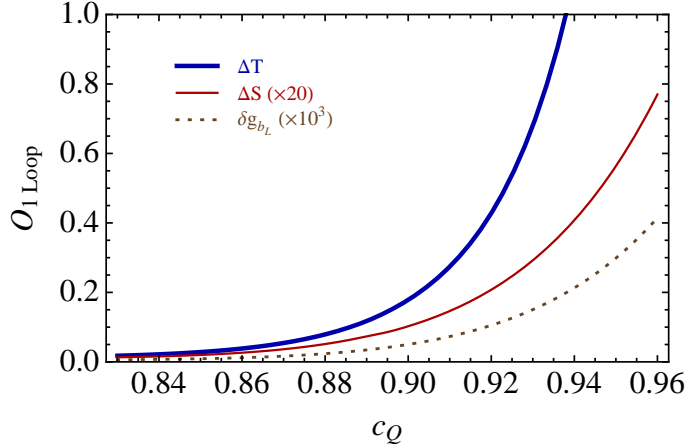}
} 
\caption{One-loop contribution to $T$, $S$ and $\delta g_{b_{L}}$,
as a function of $c_{Q}$, in scenario 2c (see text).  We take $\nu =
0.4$, $kL_{1} = 0.2$ and $\tilde{k}_{\rm eff} = 1~{\rm TeV}$ (which
corresponds to $M_{\rm KK} \approx 2.3~{\rm TeV}$).}
\label{oneloop}
\end{figure}
As mentioned in Subsection~\ref{Calculability}, the 1-loop
contributions to $T$, $S$ and $\delta g_{b_{L}}$ are finite (while at
two-loop order, they are logarithmically divergent).  We have computed
these one-loop effects using the methods described
in~\cite{Carena:2006bn}, and show a numerical example in
Fig.~\ref{oneloop}.  Here we have taken a gravitational background
with $\nu = 0.4$ and $kL_{1} = 0.2$, and we have assumed scenario~$2c$
defined above, with $\tilde{k}_{\rm eff} = 1~{\rm TeV}$.  Recall that
scenarios 2 are such that the 5D Yukawa coupling is the minimal one
that still allows to reproduce the top quark mass, for given $c_{Q}$.
In this sense, these scenarios minimize the size of these one-loop
effects, which are controlled by this coupling.  We see in the figure
that the 1-loop contribution to the $T$ parameter can be significant,
and imposes an upper bound on $c_{Q}$ that is stronger than the one
coming from the top mass itself.  By contrast, the 1-loop
contributions to $S$ and $\delta g_{b_{L}}$ are relatively small (in
the figure we show $20 \times S$ and $10^{3} \times \delta
g_{b_{L}}$).  This constraint is in tension with the one due to the
(tree-level) modification of the $Zb_{L}\bar{b}_{L}$ coupling.
\begin{figure}[t]
\centerline
{
\includegraphics[width=0.49\textwidth,clip=true]{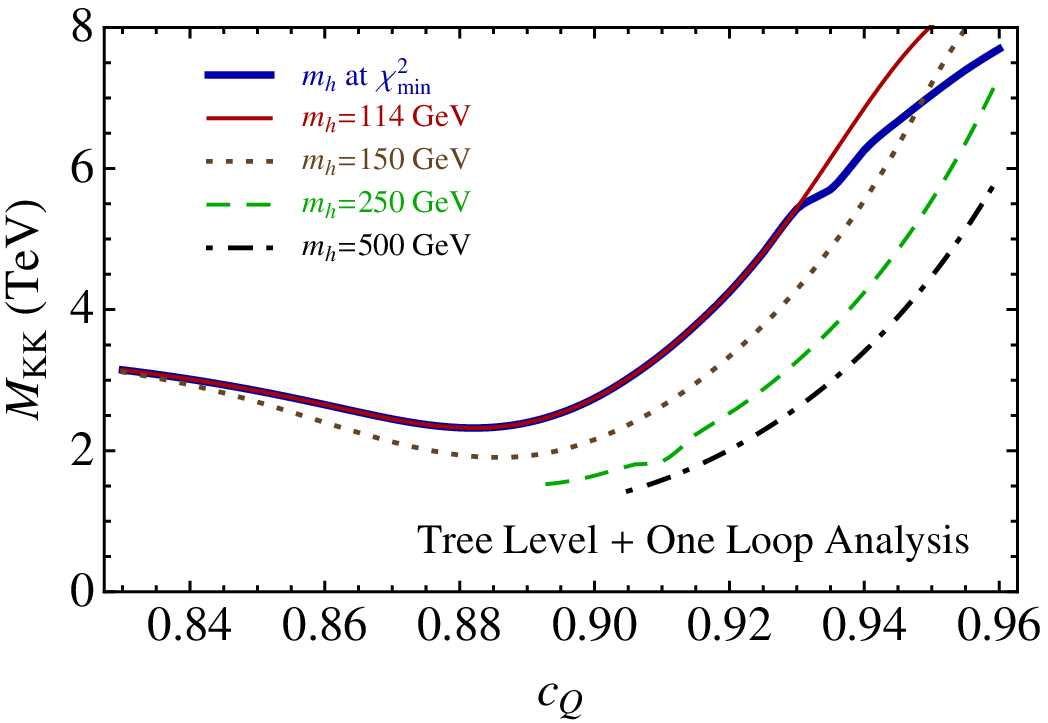}
\hspace{3mm}
\includegraphics[width=0.49\textwidth,clip=true]{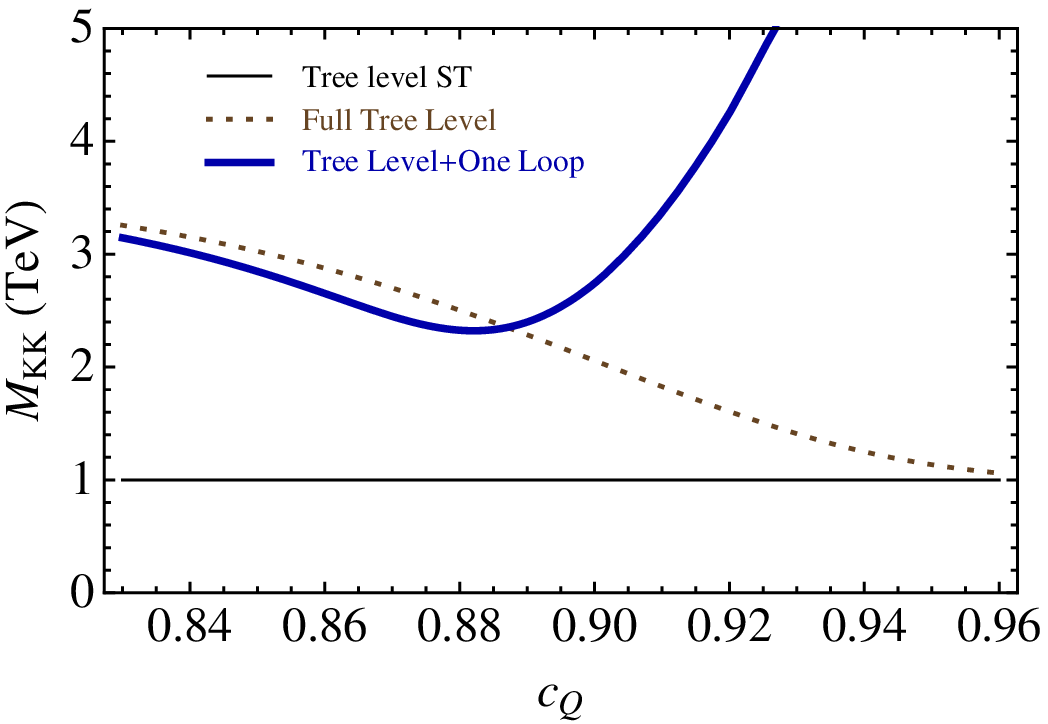}
} 
\caption{Left panel: $95\%$ CL lower bound on the mass of the first
gauge KK mode as a function of the localization of the LH top/bottom
multiplet ($c_Q$) in scenario $2c$, including one-loop effects.  The
different lines correspond to different values of $m_{h}$, with the
one marked as ``$m_{h}~\textrm{at}~\chi^{2}_{\rm min}$'' corresponding
to marginalization over $m_{h}$.  The curves are terminated where the
goodness-of-fit gives a 5\% likelihood.  Right panel: comparison of
the tree-level oblique, full tree-level, and tree-level plus 1-loop
bounds on $M_{\rm KK}$, assuming marginalization over $m_{h}$.  In all
cases, we fix $\nu=0.4$ and $kL_1=0.2$.}
\label{fig:summary}
\end{figure}
We show in the left panel of Fig.~\ref{fig:summary} the resulting
lower bound on the KK-gluon mass corresponding to \textbf{scenario
2c}.  The thick solid (blue) line corresponds to the fit procedure
used at tree-level, i.e.~evaluating $\Delta \chi^{2} = 3.84$ with a
Higgs mass that minimizes the total $\chi^{2}$ (typically near the LEP
Higgs bound).  However, we also show the bounds on the KK scale
assuming other fixed values of the Higgs mass (as would be appropriate
if a Higgs of such a mass was actually discovered).  We see from the
thick solid blue curve that --marginalizing over $m_{h}$-- a lower
95\% CL bound of $M_{\rm KK} \approx 2.3~{\rm TeV}$ is found for
$c_{Q} \approx 0.88$.  We note that for this $c_{Q}$ one has
$\chi^2_{\rm min}/{\rm dof} = 25.6/24$ at $(m_{h}, M_{\rm KK}) =
(114~{\rm GeV}, 4.3~{\rm TeV})$, which gives a goodness-of-fit with
37\% likelihood.  This results mainly from a compromise between the
(tree-level) $\delta g_{b_{L}}$ and the 1-loop contribution to $T$.
However, the black dot-dashed line shows that for a heavier Higgs,
with fixed $m_{h} = 500~{\rm GeV}$, a KK-gluon as light as $1.5~{\rm
TeV}$ ($1.4~{\rm TeV}$) would be possible, which corresponds to the
95\% CL contour about a best fit point with $\chi^2_{\rm min}/{\rm
dof} = 34/25$ ($\chi^2_{\rm min}/{\rm dof} = 37.6/25$) leading to a
10\% (5\%) likelihood.  Thus, it may be possible to have warped models
with KK-gluons around $1.5~{\rm TeV}$ that fit the EW data reasonably
well.  As a summary plot that highlights the impact of the various
contributions to the EW observables discussed above, we show in the
right panel of Fig.~\ref{fig:summary} the bounds on the KK-gluon mass,
as a function of $c_{Q}$, for the EWPT fits at a)~tree-level in the
oblique approximation [thin solid black line], b)~full tree-level
[dotted brown line] and c)~tree level plus 1-loop [thick solid blue
line].  In all cases we marginalize over $m_{h}$, although as just
pointed out this may lead to an overly pessimistic conclusion in
regards to how low $M_{\rm KK}$ could actually be.  At any rate, it is
clear from this figure that both the corrections to the
$Zb_{L}\bar{b}_{L}$ coupling and the one-loop effects play an
important role in determining the allowed $M_{\rm KK}$.

We also point out that the models consistent with the EWPT up to
1-loop order, which have a relatively low $M_{\rm KK}$, always have
KK-fermion Yukawa couplings that are \textit{perturbative}.  For
instance, at the minimum of the solid thick blue curve in
Fig.~\ref{fig:summary}, with $M_{\rm KK} \approx 2.3~{\rm TeV}$, the
4D Yukawa couplings of the form $h \, \bar{Q}_{L}^{(n)} t_{R}^{(n)}$
are all ${\cal O}(1)$, while the off-diagonal ones (i.e.~coupling
different KK levels) are much less than one (becoming smaller the
further apart the masses of the two KK modes).\footnote{By contrast,
scenarios 1 have diagonal 4D Yukawa couplings for the KK fermions of
order $3-4$, and always lead to a very large 1-loop contribution to
$T$, unless $M_{\rm KK}$ is above ${\cal O}(10~{\rm TeV})$.  In such
cases, higher-order contributions associated with the Yukawa
interactions may not be suppressed, and can have a large impact on the
EW analysis.  Nevertheless, barring tuned cancellation between these
and the UV contributions, one expects that the KK resonances will be
out of the LHC reach in such scenarios.} This is a result of the
suppression in overlap integrals associated with the non-trivial
profile of the Higgs field, much as in the KK-gauge/Higgs couplings
illustrated in Fig.~\ref{localization:fig}.  Thus, higher order
(divergent) effects involving additional powers of Yukawa couplings
are expected to be suppressed.  The most important effects that remain
are associated with QCD higher-order corrections, as mentioned in
Subsection~\ref{Calculability}.  Thus, the above results should be
taken as an illustration of how light the KK resonances can reasonably
be, as far as the EW precision constraints are concerned.

\begin{figure}[t]
\centerline
{
\includegraphics[width=0.48\textwidth,clip=true]{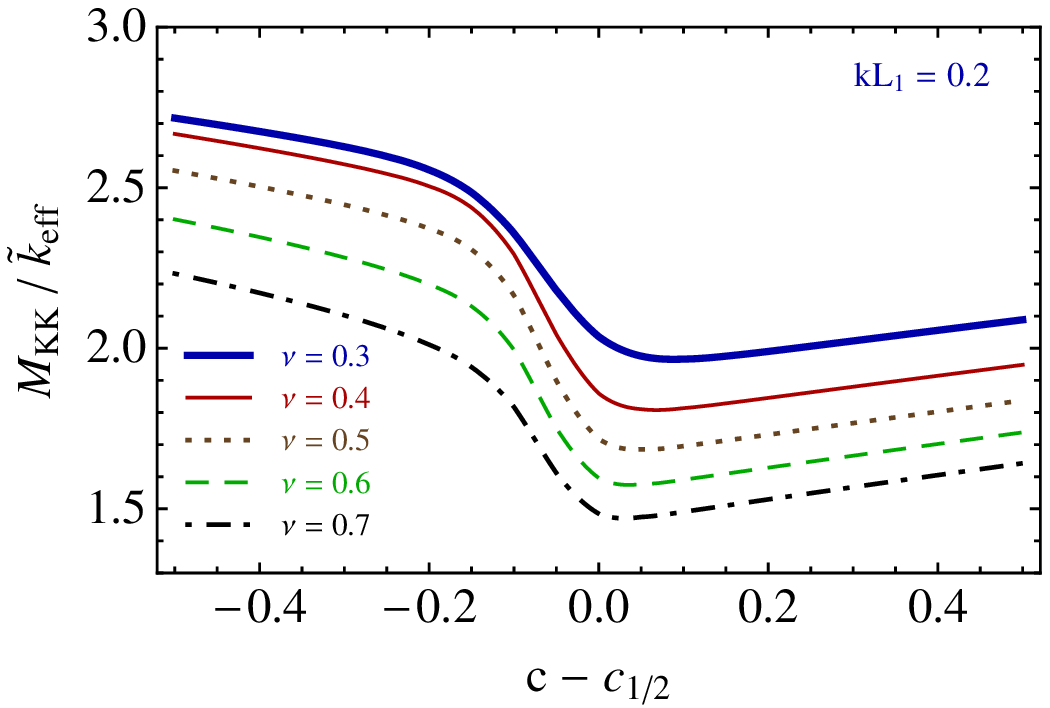}
\hspace{0.5cm}
\includegraphics[width=0.48\textwidth,clip=true]{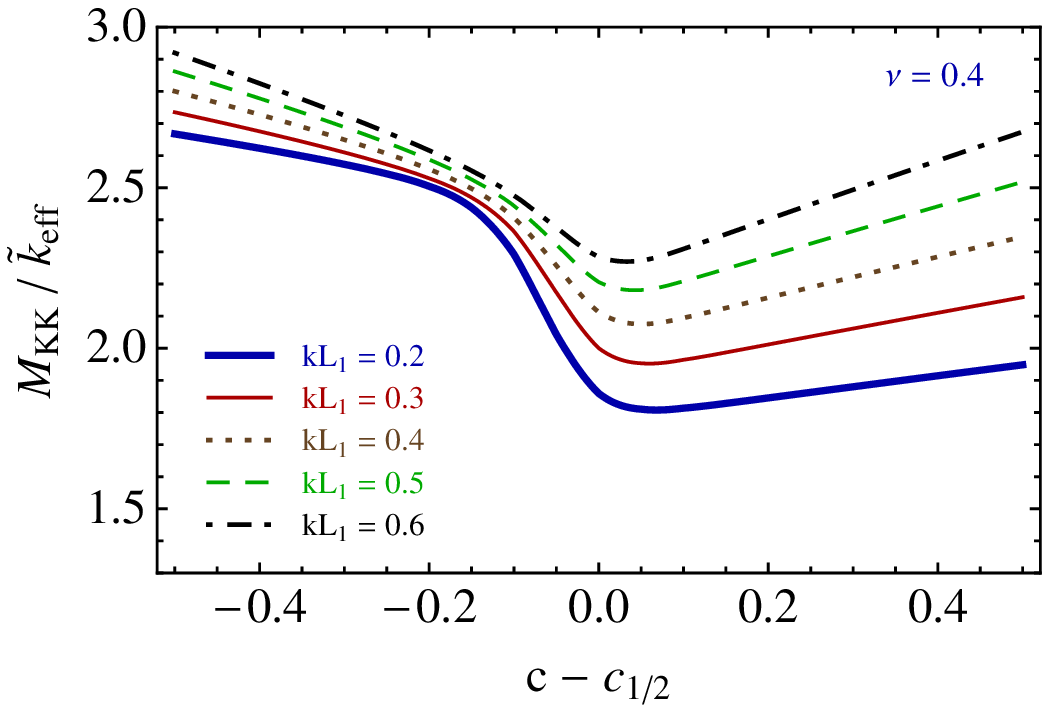}
} 
\caption{Mass of the first fermion KK mode for a $[++]$ field in
units of $\tilde{k}_{\rm eff}$ as a function of its localization
parameter (with respect to the mostly delocalized value) for values of
$\nu=0.3$, $0.4$, $0.5$, $0.6$, $0.7$ (top to bottom, left panel) and
$kL_1=0.2$, $0.3$, $0.4$, $0.5$, $0.6$ (bottom to top, right panel).
 \label{MKK:fermions}}
\end{figure}
\begin{figure}[t]
\centerline
{
\includegraphics[width=0.57\textwidth,clip=true]{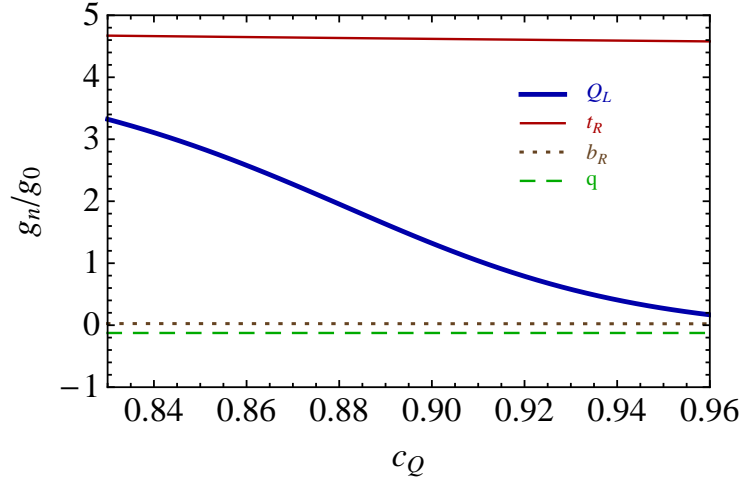}
} 
\caption{ Couplings to the first gauge KK mode (in units of $g_{0} =
g_5/\sqrt{y_1}$) of the third generation LH quarks $Q_L$, RH top
$t_R$, RH bottom $b_R$, and light quarks $q$, as a function of $c_Q$,
for scenario 2 (sub-scenarios $a$, $b$ and $c$ give similar results
for the couplings).  We fix $\nu=0.4$ and $kL_1=0.2$.  The
localization parameters $c_{T}$, $c_{B}$ and the top and bottom masses
have been fixed as described in the text.  The $b_R$ and $q$ couplings
are almost identical.}
\label{couplings:sample}
\end{figure}

These results have important implications for collider searches.
First, the absence of custodial symmetry implies a quite minimal
spectrum of massive modes.  In particular, very light
fermions~\cite{Cacciapaglia:2006gp,Contino:2008hi}, natural in
custodial models are not expected in the model under study.  We show
in Fig.~\ref{MKK:fermions} the mass of the first fermion KK mode of a
$[++]$ field as a function of the localization parameter $c$ for
different values of $\nu$ and $kL_1$.  These figures, together with
Fig.~\ref{Mkkbyktilde:gauge} show that, in some cases, the first KK
gluon can decay into a top (or bottom) and a heavy fermion, although
the latter are in this case much closer to threshold than in custodial
models.  Channels with a heavy fermion of the first and second
generations are also typically open: under the anarchic assumption
such decay channels are very suppressed, but they could be more
important in other scenarios for flavor.  Second, masses lighter than
previously considered in models with warped extra dimensions (with
semi-anarchic Yukawas) may be allowed by EWPT. This result is very
sensitive to the localization of the third generation quarks which
determine the coupling of the top and bottom quarks to the gauge KK
modes, as well as the KK-fermion Yukawa couplings.  We show in
Fig.~\ref{couplings:sample} the couplings of the different quarks to
the first KK gluon in units of $g_{0} = g_5/\sqrt{y_1}$ for
scenarios~$2$, with $\nu=0.4$ and $kL_1=0.2$.  We see that the largest
is the coupling to $t_{R} \bar{t}_{R}$, followed by $Q_{L}
\bar{Q}_{L}$ and then to light quark pairs.  Thus, the KK-gluon decays
dominantly into RH top pairs (or, perhaps, a channel involving one
KK-fermion and the associated zero-mode), bearing some resemblance to
the scenario of Eq.~(\ref{couplings:traditional}).  However, the
reduced couplings to the light quarks can make the discovery more
challenging, for a given $M_{\rm KK}$.
%

\section{Phenomenological Implications}
\label{collider}

In this section we present the main collider implications of some
selected points in parameter space.  As we have seen in the previous
section, the EW fit prefers models of type $\bf 2$, which always have
the KK-gluons relatively strongly coupled to $t_{R}$ (see
Figs.~\ref{fig:summary} and \ref{couplings:sample}).  Thus, the
$t\bar{t}$ decay channel for the KK-gluon is always dominant.  To
illustrate the expected signal at the LHC, we consider two models, as
suggested by the analysis in the previous section.

The first one corresponds to scenario ${\bf 2c}$ with $\nu=0.4$,
$kL_1=0.2$ and $c_Q=0.88$.  With $m_{h} = 114~{\rm GeV}$, we find a
$\chi^{2}/{\rm dof} = 25.6/24$, giving a likelihood of $37 \%$.  This
fit is only slightly worse than the SM one: the total $\chi^{2}$ is
slightly reduced, but there is one additional degree of freedom,
corresponding to $\tilde{k}_{\rm eff}$.  The KK gluon mass is
\begin{equation}
M_{\rm KK} \approx 2.3~\mathrm{TeV}
\hspace{1cm}
\textrm{(95\% CL)}~.
\end{equation}
We also find that the first KK resonance of the third generation quark
$SU(2)_{L}$ doublet has a mass $M_{Q} \approx 2.1~{\rm TeV}$, while
the first KK resonance of the bottom $SU(2)_{L}$ singlet has a mass
$M_{B} \approx 1.85~{\rm TeV}$.  Both are sufficiently light for the
decays $G^{(1)} \to Q^{(0)}_{L} \bar{Q}^{(1)}_{L}$ and $G^{(1)} \to
b^{(0)}_{R} \bar{B}^{(1)}_{R}$ to be open.  The corresponding
couplings to fermion pairs are
\begin{eqnarray}
&g_{Q_L}\approx 1.95 g_s~, \quad
g_{t_R}\approx 4.63 g_s~, \quad
g_{b_R}\approx 0.02 g_s~, \quad
g_{q}\approx -0.13 g_s~,
\label{couplings21}
\\ [0.5em]
&g_{q_L,Q^{(1)}_L}\approx 3.85 g_s~, \quad
g_{b_R,B^{(1)}_R}\approx 1.15 g_s~,
\label{couplings22}
\end{eqnarray}
where we omitted the superscripts for the zero modes, and the first
line refers to SM fermion pairs.  The first KK-resonance of the top
$SU(2)_{L}$ singlet is heavier than the KK-gluon, so that this channel
is kinematically closed.  However, the first KK resonances of all the
remaining SM fermions are lighter than the KK-gluon, and are therefore
open as decays of the form $G^{(1)} \to q^{(0)} q^{(1)}$.
Nevertheless, the relevant couplings are all significantly smaller
than $g_{s}$, so that these are somewhat rare decays that, in spite of
the multiplicity, do not change appreciably the width of the KK-gluon.
With the above couplings, we find that $\Gamma_{G^{(1)}} \approx 710$
GeV, and that the KK-gluon has the following branching fractions:
\begin{eqnarray}
& BR(G^{(1)}\to t\bar{t})\approx 0.81~,
\quad
BR(G^{(1)}\to b\bar{b}) \approx 0.12~,
\quad
BR(G^{(1)}\to q\bar{q})\approx 0.004~,\\ [0.5em]
& BR(G^{(1)}\to t^{(1)}_{L} t)\approx 0.02~,
\quad
BR(G^{(1)}\to b^{(1)}_{L} b)\approx 0.04~,
\quad
BR(G^{(1)}\to b^{(1)}_{R} b) \approx 0.01~,
\end{eqnarray}
where in the last line both conjugate processes (e.g. $G^{(1)}\to
t^{(1)}_{L} \bar{t}$ and $G^{(1)}\to \bar{t}^{(1)}_{L} t$) are
understood.  The total $G^{(1)}$ production cross section is $\sim
24~{\rm fb}$ ($\sim 188~{\rm fb}$) at a CM energy of $7~{\rm TeV}$
($14~{\rm TeV}$).

The second model has instead $c_Q=0.906$ which, with $m_{h} = 500~{\rm
GeV}$, leads to $\chi^{2}/{\rm dof} = 34/25$, giving a likelihood of
$10 \%$ (still reasonably large).  The KK-gluon mass is now
\begin{equation}
M_{\rm KK} \approx 1.5~\mathrm{TeV}~
\hspace{1cm}
\textrm{(95\% CL)}~,
\end{equation}
while the first KK resonance of the third generation quark $SU(2)_{L}$
doublet has a mass $M_{Q} \approx 1.3~{\rm TeV}$, and the first KK
resonance of the bottom $SU(2)_{L}$ singlet has a mass $M_{B} \approx
1.2~{\rm TeV}$.  The corresponding couplings to fermion pairs are
\begin{eqnarray}
&g_{Q_L} \approx 1.17~, \quad
g_{t_R} \approx 4.62~, \quad
g_{b_R} \approx -0.03 g_s~, \quad
g_{q} \approx -0.13 g_s~,
\label{couplings11}
\\ [0.5em]
&g_{q_L,Q^{(1)}_L}\approx 3.16 g_s~, \quad
g_{b_R,B^{(1)}_R}\approx 1.15 g_s~.
\label{couplings12}
\end{eqnarray}
Now we have $\Gamma_{G^{(1)}} \approx 390$ GeV and the following
branching fractions:
\begin{eqnarray}
& BR(G^{(1)}\to t\bar{t})=0.83~,
\quad
BR(G^{(1)}\to b\bar{b})=0.05~,
\quad
BR(G^{(1)}\to q\bar{q})=0.005~,\\ [0.5em]
& BR(G^{(1)}\to t^{(1)}_{L} t)\approx 0.03~,
\quad
BR(G^{(1)}\to b^{(1)}_{L} b)\approx 0.06~,
\quad
BR(G^{(1)}\to b^{(1)}_{R} b) \approx 0.02~.
\end{eqnarray}
The total $G^{(1)}$
production cross section is $\sim 0.19~{\rm pb}$ ($\sim 1.3~{\rm
pb}$) at a CM energy of $7~{\rm TeV}$ ($14~{\rm TeV}$).  

The BR's in these models are relatively similar to the benchmark of
Ref.~\cite{Agashe:2006hk}, but with lighter masses and reduced
couplings to the light quarks.  Note, however, that the first model
has a non-negligible branching fraction into bottom pairs, and that
there are ``exotic'' channels involving a KK fermion with BRs at the
few percent level (in the case of the third generation; for the first
two generations the corresponding BRs are expected to be much smaller,
although the precise values depend on the details of how flavor is
implemented).

\begin{figure}[t]
\centerline
{
\includegraphics[width=0.5\textwidth,clip=true]{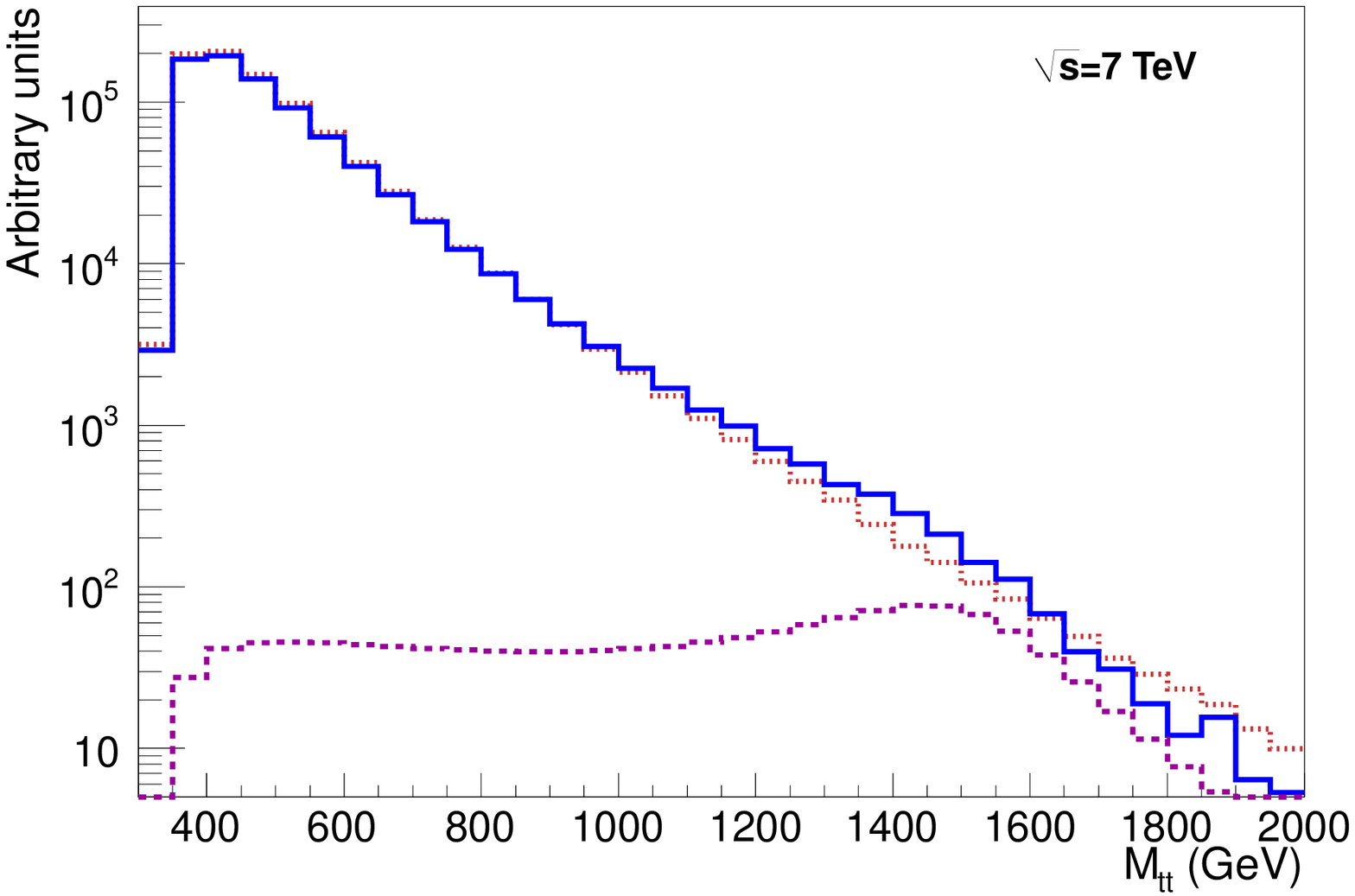}
\includegraphics[width=0.5\textwidth,clip=true]{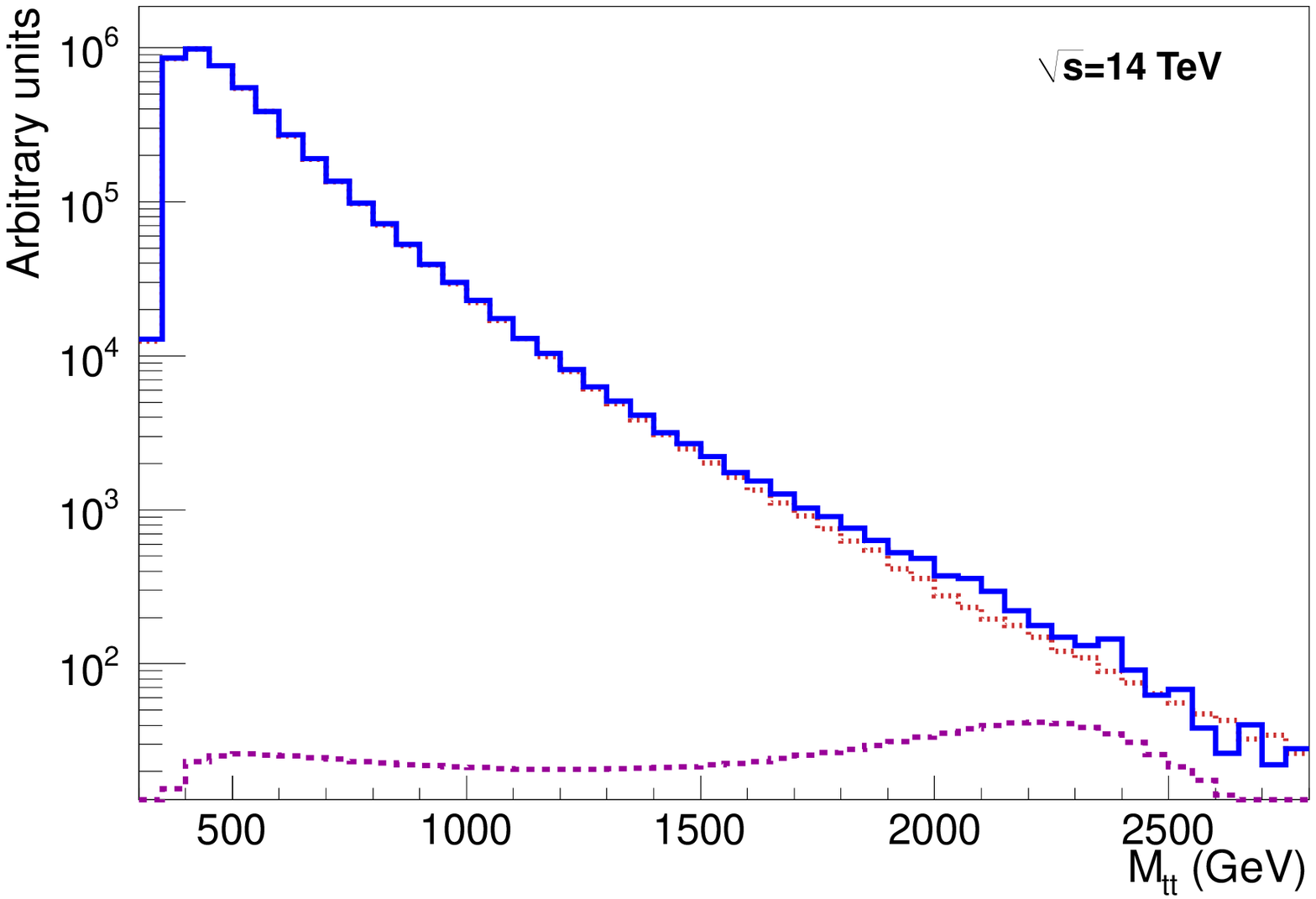}  
} 
\caption{$t\bar{t}$ invariant mass distribution in the SM (red
dotted), in the model with extra-dimensional physics (solid blue) and
the contribution of just the KK gluon exchange (dashed purple).  The
left panel corresponds to a KK-gluon with $M_{\rm KK} \approx
1.5~\mathrm{TeV}$ and the couplings of
Eqs.~(\ref{couplings11})-(\ref{couplings12}).  The right panel
corresponds to $M_{\rm KK} \approx 2.3~\mathrm{TeV}$ and the couplings
of Eqs.~(\ref{couplings21})-(\ref{couplings22}).}
\label{ttbar:model1}
\end{figure}
We have implemented these models in MADGRAPH/MADEVENT
v4~\cite{Alwall:2007st}, using PYTHIA 6~\cite{pythia} for
hadronization and showering and PGS4~\cite{PGS4} for detector
simulation.  We use the \textbf{CTEQ6L1} parton distribution function,
with the QCD renormalization and factorization scales equal to the
central $m_T^2$ of the event.  We have then implemented the $t\bar{t}$
CMS analysis of Ref.~\cite{ttbar:analysis}, which we use to normalize
our SM sample, generated by MG/ME, once we check that the shape of the
$t\bar{t}$ invariant mass distribution is reasonably reproduced.  We
use this (re)normalization factor for the signal $+$ background,
taking into account interference effects.  We consider here the
semi-leptonic $t\bar{t}$ channel, separating the $e$ and $\mu$
channels~\cite{ttbar:analysis}.  We show in Fig.~\ref{ttbar:model1}
the expected $t\bar{t}$ invariant mass distribution at the LHC, at the
partonic level.  The left (right) panel corresponds to the second
(first) model above at $\sqrt{s}=7$ TeV ($\sqrt{s}=14$ TeV).  In both
plots we represent the SM prediction with a red dotted line and the
prediction of the model (including the interference with the SM
$t\bar{t}$ contribution) in solid blue.  Also, just to guide the eye,
we show the contribution assuming only the KK gluon exchange as a
purple dashed line.  Although the lightest mass case shows a slight
excess over background, these results suggest that extracting the
signal will be challenging.

\section{Conclusions}
\label{conclusions}

We have studied a recently proposed model in five dimensions with a
warped background~\cite{Cabrer:2010si} that can be consistent --from
the point of view of EW precision constraints-- with new gauge boson
resonances at about $1.5~{\rm TeV}$.  This represents a significant
reduction of the bound on the new physics scale compared to fairly
elaborate models based on AdS$_{5}$ backgrounds.  The scenario is also
relatively simple, with just the SM field content promoted to 5D, plus
a new ``stabilizing'' scalar field that is responsible for the
all-important deviation from AdS$_{5}$ near the IR brane.  A simple
way to understand the relaxation of the bound can be obtained by
considering the localization of the relevant fields: the gauge boson
KK modes are strongly attracted towards the IR brane, while it is
still possible to accommodate a profile for the Higgs field that,
although mostly IR localized (so that the RS solution to the hierarchy
problem is preserved), attains its maximum before reaching the IR
brane.  As a result, any overlap integral that involves the Higgs and
the (first) gauge or fermion KK modes can be dramatically reduced
compared to cases where all the fields are localized within a distance
of order $1/k$ from the IR brane, as happens in the AdS$_{5}$
background.  Since the deviations from the SM are determined by such
overlap integrals (and the scale of the new resonances), lighter
states can be allowed.

Here we have extended the study in Ref.~\cite{Cabrer:2010si} to the
case where the SM fermions can be arbitrarily localized along the
extra-dimension, which allows to understand the SM flavor structure as
arising from the localization of fermion fields.  We point out that
reproducing the observed top quark mass, which requires the top quark
to be localized sufficiently close to the IR brane, can result in
significant constraints from anomalous contributions to the
$Zb_{L}\bar{b}_{L}$ coupling, beyond the constraints from the oblique
parameters considered in Ref.~\cite{Cabrer:2010si}.  This is a direct
consequence of the absence of any custodial
protection~\cite{Agashe:2003zs} together with the ``anarchical''
explanation of flavor.  In addition, we find that the 1-loop
corrections to the EW observables, which are finite in these
scenarios, strongly restrict the allowed region of parameter space.
In particular, a tension between the (tree-level) corrections to the
$Zb_{L}\bar{b}_{L}$ coupling and the 1-loop contribution to the
Peskin-Takeuchi $T$-parameter, strongly constrains the localization of
the third generation quarks and the resulting 95\% CL bound on the KK
scale.  Nevertheless we find that for an optimal top/bottom
$SU(2)_{L}$ doublet localization, and when the 5D bottom Yukawa
coupling is a factor of a \textit{few} smaller than the top one, the
KK-gluons can be as light as $1.5~{\rm TeV}$ if the Higgs is heavy,
with the likelihood inferred from the EW fit still being reasonably
large.  However, if the Higgs mass is allowed to float in the fit, one
finds a 95\% CL lower bound of $M_{\rm KK} \approx 2.3~{\rm TeV}$,

A KK-gluon mass of about $1.5~{\rm TeV}$ can open the exciting
possibility that the Randall-Sundrum solution to the hierarchy
problem, based on warped compactifications, and with a bulk SM field
content, could lead to observable resonances at the LHC with lower
luminosities than previously thought.  Compared to the widely studied
AdS$_{5}$ framework, the result arises from two opposing effects: the
previously mentioned strong localization of the gauge boson resonances
towards the IR brane implies a reduction of its couplings to the light
fermions, thus leading to a suppression in production.  This can be
compensated by the lower allowed mass of the gauge KK modes.  However,
we find that discovering such a resonance in the dominant $t\bar{t}$
channel is likely to be challenging.  Boosted top techniques and a
very detailed knowledge of the $t\bar{t}$ tail will likely be required
to discover these modes.  We also find that, unlike in the case of an
AdS$_{5}$ background without the custodial symmetry, the present class
of models allows for the production of single KK-fermion resonances of
the SM fields, in KK-gluon decays.  Although these decay channels are
likely subdominant (in particular, they do not dramatically change the
KK-gluon width), they may provide an interesting handle on the
fermionic resonances.\footnote{Note that in models with custodial
symmetry, fermion custodians are expected to be in general lighter and
these channels are likely to be more
relevant~\cite{Carena:2007tn,Vignaroli:2011ik}.  The discovery of new
fermions with exotic charges, like
$Q=5/3$~\cite{Cacciapaglia:2006gp,Contino:2008hi}, would in any case
be a clear smoking gun of custodial models.} Nevertheless, rather high
luminosities are likely necessary.  One should recall, however, that
we are working under the (semi-)anarchic assumption of flavor.  If the
light families are closer to the IR brane (with a correspondingly
smaller 5D Yukawa coupling) the production cross section can be
larger, and relatively light spin-1 resonances in the 1-2 TeV range
could be more readily observed at the LHC.

\section*{Acknowledgments}

We would like to thank Kaustubh Agashe, Mert Aybat, 
C\'edric Delaunay, Gero von Gersdorff and especially Manuel P\'erez-Victoria
for useful discussions.  A.C. and J.S. are supported by MICINN
(FPA2006-05294,FPA2010-17915, FPU and Ram\'on y Cajal programs) and
Junta de Andaluc\'{\i}a (FQM 101, FQM 03048 and FQM-6552).  E.P. is
supported by DOE grant DE-FG02-92ER40699.


\end{document}